\documentclass[twocolumn,showpacs,preprintnumbers,amsmath,amssymb]{revtex4}

\usepackage{graphicx}
\usepackage{dcolumn}
\usepackage{bm}

\begin{document}

\preprint{APS/123-QED}

\title{High Resolution Search for Exotic Pentaquark $\Theta^{++}$, 
and $\Theta^+$ at Jefferson Lab }

\author{Haiyan Gao and Wang Xu}
\affiliation{Department of Physics\\ Duke University and the Triangle 
Universities Nuclear Laboratory\\}

\date{\today}

\begin{abstract}
Recent evidence for the existence of pentaquark $\Theta^+$ particle from
several experiments at several different laboratories around the world 
has caused great excitement and raised many unanswered questions.
We discuss a new, high resolution experiment searching for pentaquark 
states in Hall C at Jefferson Lab using an untagged bremsstrahlung photon beam 
employing both the hydrogen and the deuterium targets by 
studying the following processes: 
$\gamma p \rightarrow \Theta^{++} K^-$, 
and $\gamma n \rightarrow 
\Theta^+ K^-$. 
This new experiment will significantly improve our current 
knowledge of the mass, 
the width of the $\Theta^+$ particle if it is confirmed 
and provide unambiguous evidence for the
 existence or non-existence of the $\Theta^{++}$ particle from the 
$\gamma p \rightarrow \Theta^{++} K^-$ process.

\end{abstract}

\maketitle

\section{Introduction and Physics Motivation}

The first evidence of the observation of a pentaquark $\Theta^+$ 
particle came in early 2003
from the LEPS collaboration~\cite{leps} in which 
a sharp resonance was reported at a mass of $(1.54 \pm 0.01)$ (GeV)/c$^2$.
The DIANA Collaboration~\cite{diana} reported evidence for a resonance 
enhancement at $M=1539 \pm 2$ MeV/c$^2$ and $\Gamma \le 9$ MeV/c$^2$ in the
charge-exchange reaction $K^+Xe \rightarrow K^0 p Xe'$. The 
CLAS collaboration~\cite{clas}
reported a statistical significance of $(5.3 \pm 0.5) \sigma$ at 
the $K^+n$ invariant mass peak of $1542 \pm 5$ MeV/c$^2$ with a measured width 
of 21 MeV for the reaction $\gamma d \rightarrow K^+ K^- p(n)$. 
The SAPHIR~\cite{saphir} collaboration reported
a peak at $M_{\Theta^+}=1540 \pm 4 \pm 2$ MeV (4.8 $\sigma$ confidence level) 
and an upper limit of 25 MeV for
the width with a $90\%$ confidence level in the $K^+ n$ invariant mass 
distribution from the $\gamma p \rightarrow nK^+K^0_s$ process.
The SAPHIR Collaboration also reported the absence of a signal in
the $K^+p$ invariant mass distribution in the $\gamma p \rightarrow p K^+K^-$
reaction.
More recently, the HERMES collaboration~\cite{hermes} reported evidence 
of the $\Theta^+$ particle from a deuterium target with the decay mode 
$\Theta^+ \rightarrow p K^0_s \rightarrow p \pi^+\pi^-$ at $1526 \pm 2 \pm 2$ 
(MeV) with a width of 7.5 MeV dominated by detector resolution. 
The CLAS Collaboration~\cite{clas2} 
also reported evidence on $\Theta^+$ from a hydrogen target from the 
$\gamma p \rightarrow \pi^+ K^- \Theta^+$ process. 
They found a peak in the invariant mass of $nK^+$ at 1555 MeV with 
a width of about 26 MeV. 

The CERN NA49 Collaboration~\cite{NA49} reported an 
exotic $S=-2, Q=-2$ baryon resonance 
in proton-proton collisions with a mass of $1862 \pm 2$ MeV/c$^2$ and width 
below the detector resolution of 18 MeV/c$^2$ in the $\Xi^-\pi^-$, 
and  $\Xi^-\pi^+$
invariant mass spectra. These two states are believed to be candidates for 
the $dsds\bar{u}$ ($\Xi^{--}_{\frac{3}{2}}$), and the 
$dsus\bar{d}$ ($\Xi^{0}_{\frac{3}{2}}$) pentaquark state, respectively.
However, there is probably more evidence against the existence
of the $dsds\bar{u}$ and $dsus\bar{d}$ pentaquark states ~\cite{negative2}.
Mostly recently, the H1 collaboration reported~\cite{H1} evidence for a narrow 
anti-charmed baryon state, interpreting such a state as the pentaquark
$uudd\bar{c}$ state.

In summary,  the experimental situation concerning the $\Theta^+$ and 
$\Theta^{++}$ pentaquark states is the following.
There is evidence from different experiments supporting the existence 
of the pentaquark $\Theta^+$ state, but there is no evidence for 
the $\Theta^{++}$ particle. The mass of the $\Theta^+$ particle is from 
1527 MeV to 1555 MeV, and the width is found to be from 9 MeV to 26 MeV 
limited by detector resolutions. 
Also, the existing 
data seem to suggest that the $\Theta^+$ particle is an isoscalar particle.
However, the existence of $\Theta^{+}$ pentaquark is far from being confirmed
experimentally~\cite{negative}.
One also needs to address possible experimental issues
such as kinematic reflection~\cite{dzierba} due to 
the decay of higher mass mesons, such as the $f_2(1275)$, the $a_2(1320)$, 
and the $\rho_3(1690)$.

The original chiral soliton model~\cite{soliton} predicted  
a mass of 1530 MeV and a total width of less than 15 MeV for the 
$\Theta^+$ particle. It also predicted that the $\Theta^+$ particle is
a spin $\frac{1}{2}$ isoscalar particle and the predicted mass 
for the $\Xi_{3/2}$ state is 2070 MeV. 
In the correlated diquark picture by Jaffe and Wilczek~\cite{jaffe},
the diquarks QQ are correlated in an antisymmetric color, 
flavor and spin state.
In the case of the $\Theta^+$ pentaquark state, it is a bound state of two
highly correlated $ud$ pairs and an antiquark. 
The narrowness of the width of the $\Theta^+$ particle 
can be explained in this model by the weak coupling between the $K^+n$ 
continuum and the bound state of the two diquark pairs and the antiquark, 
$[ud]^{2}\bar{s}$. One of the interesting predictions from this model,
 which 
differs from the chiral soliton model prediction, is the mass of the cascade 
isospin $\frac{3}{2}$ multiple $\Xi_{3/2}([us]^2\bar{d})$. It is 
predicted to be around 1750 MeV 
with a width about $50\%$ greater than that of the $\Theta^+$ particle. 

 The $\Theta^+$ particle was also hypothesized as
an isotensor resonance to explain the observed narrow width of 
the $\Theta^+$ particle~\cite{cpr} 
via isospin-violating strong decays. 
The hypothesis that $\Theta^+$ is an isotensor particle 
implies the existence of $\Theta^{**+++}$, $\Theta^{**++}$, $\Theta^{**0}$, and
$\Theta^{**-}$ states. 
Gerasyuta and Kochkin~\cite{gk} calculated the mass spectra of the isotensor
Theta-pentaquarks with $J^{p} ={\frac{1}{2}}^{\pm}, {\frac{3}{2}}^{\pm}$ in a
relativistic quark model. The predicted mass for $\Theta^{**++}$ is
1575 MeV (1761 MeV) for $J^{p} ={\frac{1}{2}}^{\pm}$ 
($J^{p} ={\frac{3}{2}}^{\pm}$).
Here we following the convention in the literature that $\Theta$ represents
the I=0 state in anti-decuplet, $\Theta^{*}$ represents the I=1 state in 27-plet and
$\Theta^{**}$ the I=2 state in 35-plet. In the remaining of the paper, 
we will not explicitly differentiate the $\Theta^{**++}$ and the $\Theta^{*++}$ states
because the proposed experiment will not be able to determine 
which multiplet the observed $\Theta^{++}$ particle belongs to.

Walliser and Kopeliovich~\cite{walliser} 
investigated the implications of the $\Theta^+$ exotic pentaquark for the 
baryon spectrum in topological soliton models. They estimate the positions of
other pentaquark and septuquark states with exotic and with non-exotic
quantum numbers, particularly within the 27-plet baryon and the
35-plet multiplets in SU(3) soliton model. 	
In the 27-plet, the $J={\frac{3}{2}}, 
T=1$ multiple ($\Theta^{*0}$, $\Theta^{*+}$, $\Theta^{*++}$) are estimated 
to have a mass of 1.65 to 1.69 GeV. 
Most recently, Wu and Ma~\cite{ma_private} studied the mass and width of 
the pentaquark $\Theta^{*}$ states in the {27} baryon multiplet from chiral 
soliton model. Their calculations show that the mass of $\Theta^{*}$ is 
about 1.61 GeV and the width for the process $\Theta^{*} \rightarrow KN$ is 
less than 44 MeV.
Bijker {\it et al.}~\cite{bijker} constructed a complete classification of $qqqq\bar{q}$ 
states in terms of the flavor-spin SU(6) representation and found that only
the anti-decuplet, 27-plets and the 35-plets contain exotic states which can
not be constructed by three quarks only. In this model, the ground state 
pentaquark is identified as the observed $\Theta^+ (1540)$ state, and is 
predicted to be an isosinglet anti-decuplet state. The predicted masses 
for the excited exotic baryons are 1660 MeV and 1775 MeV.

While it is important to determine the quantum number of the $\Theta^+$ 
particle, particularly the spin and the parity and to search for other 
members in the exotic baryon family,
it is more essential to 
confirm the existence of 
the $\Theta^+$ particle with significantly improved statistics, and to 
determine its mass and the width more precisely.
Karliner and Lipkin~\cite{karliner} derived
an upper bound on the mass difference between the 
$\Xi^\ast$ and $\Theta^+$ based
on simple assumptions about $SU(3)_f$ symmetry breaking and quantum mechanical
variational method. The resulting rather robust bound is more than 20 MeV below 
the experimentally reported $\Xi^\ast$ - $\Theta^+$ mass difference. 
It is also very important to verify definitely the non-existence or the 
existence of the $\Theta^{++}$ particle and to search for the $\Theta^{*++}$ 
particle. 

\begin{figure}
\includegraphics[width=3.5in]{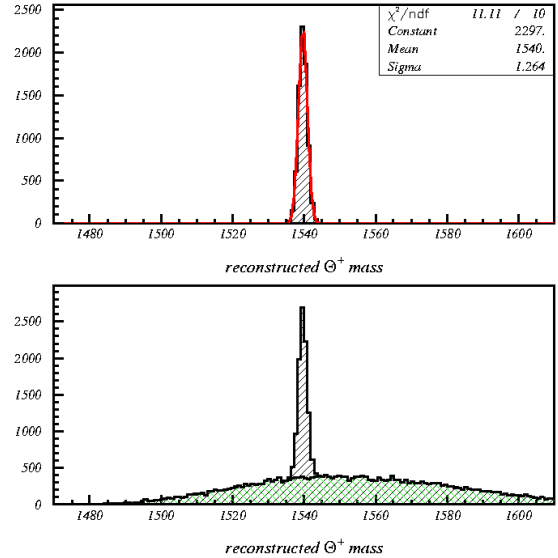}
\caption{Reconstructed $\Theta^{+}$ mass spectra for the proposed 
experiment. In this simulation, physical events were generated from the 
$\gamma n \rightarrow \Theta^{+} K^-$ process with a $\Theta^{+}$ mass of 
1540 MeV; The background is from $K^+K^-$ pairs and an accidental 
coincidental rate. }
\label{fig:thetaplus}
\end{figure}

We proposed~\cite{gao} a new, high resolution 
search for pentaquark 
$\Theta^+$, $\Theta^{++}$ and $\Theta^{*++}$ in Hall C at Jefferson Lab 
using high resolution magnetic spectrometers 
in combination with a neutron counter by studying two different 
physical processes: the 
$\gamma n \rightarrow \Theta^{+}K^-$, where $\Theta^+$ decays via 
$\Theta^+ \rightarrow K^+ n$, and the 
$\gamma p \rightarrow \Theta^{++} (\Theta^{*++}) K^-$, where $\Theta^{++} 
\rightarrow K^+ p$.
Due to the lack of a free neutron target, 
a liquid deuterium target will be used for the 
$\gamma n \rightarrow \Theta^+ K^-$ 
process. 
The $K^-$ particle will be detected in the Short Orbit Spectrometer (SOS) 
in coincidence with the $K^+$ and the neutron which will be detected 
in the High resolution Kaon Spectrometer (HKS), 
and the neutron counter, respectively. By detecting all three
particles, the untagged bremsstrahlung photon energy, the mass of the 
$\Theta^+$ particle, and the initial neutron momentum
inside the deuteron can be reconstructed completely.

For the two-body process $\gamma p \rightarrow \Theta^{++} (\Theta^{*++}) 
K^-$ process
one can not reconstruct the mass of the undetected $\Theta^{++}$ 
($\Theta^{*++}$) particle from detecting $K^-$ particle only unless the 
energy of the incident real photon is known.  
Assuming the $\Theta^{++}(\Theta^{*++})$ particle decays into  $K^+$ and $P$,
the incident real photon energy can be
reconstructed by detecting $K^+$ in coincidence with the $K^-$ particle. 
Such 
coincidence detection also suppresses 
backgrounds from other physical processes which have final states 
different from the $K^+ K^- p$ final state. 
The proposed experiment
allows for the search of the $\Theta^{++}$ ($\Theta^{*++}$) particle 
in the mass range of 1500 MeV
to 1800 MeV.

\begin{figure}
\includegraphics[width=3.5in]{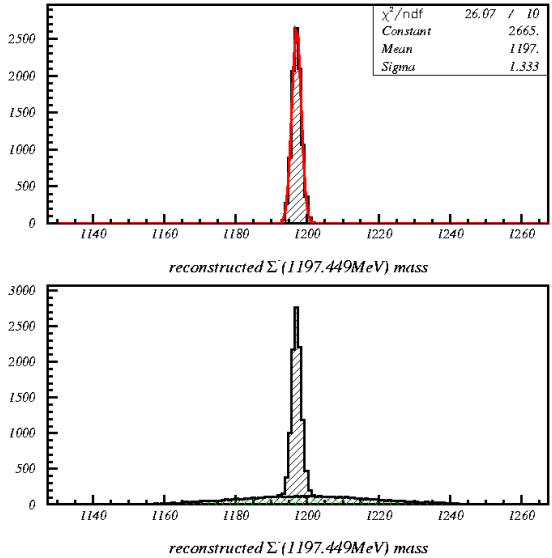}
\caption{The reconstructed $\Sigma^-$ mass from the $\gamma n \rightarrow \Sigma^- K^+$, $\Sigma^- \rightarrow \pi^- n$ reaction (see text). } 
\label{fig:calibration}
\end{figure}

The primary sources of the physics backgrounds for the proposed 
$\gamma p \rightarrow \Theta^{++} (\Theta^{*++}) K^-$ process and the
$\gamma n (D) \rightarrow \Theta^{+} K^-$ process are: 
(i) $\phi$ meson production, 
(ii) $\gamma + p \rightarrow K^{+} +\Lambda (1520)$ where, 
$\Lambda (1520) \rightarrow p + K^{-}$, and
(iii) three-body final state $K^{+}K^{-}$ production.
Monte Carlo simulations have been carried out to study these backgrounds, 
taking into account the momentum and angular resolutions of the 
two hadron spectrometers 
and their full momentum and angular acceptances.  
The spectrometer momentum and angle settings based 
on the two-body kinematics of the $\gamma p \rightarrow \Theta^{++} (\Theta^{*++}) K^-$
and the $\gamma n \rightarrow \Theta^+ K^-$
processes together with the reconstructed photon energy cut 
($E_0 - 125 \le E_{\gamma} \le E_0 - 25$ MeV), where $E_0$ is the incident
electron beam energy, effectively suppress the background contributions from
the aforementioned physical processes. Cuts can be applied in the 
invariant mass of the $K^+ p$ ($K^+ n$) system in order to suppress these 
backgrounds further. 
Simulations show that both the $\phi$ production and the $\Lambda (1520)$ 
channels are suppressed even without a 100 MeV of $E_\gamma$ cut and 
kinematic reflections are not issues for the proposed experiment.
Fig.~\ref{fig:thetaplus} shows the projected 
result on the $\Theta^+$ particle search.

The $\gamma n \rightarrow \Sigma^- K^+$ reaction will 
be used for the mass calibration. 
The $\Sigma^-$ particle will be reconstructed by its decay into 
neutron and $\pi^-$ particle. The neutron will be detected in the 
neutron counter in coincidence with the $\pi^-$, and the mass of the $\Sigma^-$
particle will be reconstructed in a similar way as that of the $\Theta^+$
particle.  
To suppress the background the $K^+$ particle will be detected in the SOS 
spectrometer in coincidence with the neutron and the $\pi^-$ particle. 
Fig.~\ref{fig:calibration} shows the simulated $\Sigma^-$ particle mass
determination, which is better than 0.3 MeV.
The upper panel shows the simulation result for the 
$\gamma n \rightarrow \Sigma^- K^+$, $\Sigma^- \rightarrow \pi^- n$ 
reaction only, while the 
lower panel result includes accidental and other possible physical 
backgrounds. 
This new experiment will determine 
the absolute mass of the $\Theta^+$ particle to better than 0.5 MeV 
and determine the $\Theta^+$ particle width to
better than 1.3 MeV ($\sigma$).

This work is supported 
by the U.S. Department of Energy under 
contract number DE-FG02-03ER41231.
The author thanks the Yukawa Institute for Theoretical Physics
at Kyoto University.
Discussions during the YITP workshop YITP-W-03-21 on
``Multi-quark Hadron: four, five and more?''
were useful for the completion of this work.



\begin{references}


\bibitem{leps}
T. Nakano {\it et al.}, Phys. Rev. Lett. {\bf 91}, 012002 (2003).

\bibitem{diana}
V.V. Barmin {\it et al.}, hep-ex/0304040.

\bibitem{clas}
S. Stepanyan {\it et al.}, hep-ex/0307018, Phys. Rev. Lett. {\bf 91}, 
252001 (2003).

\bibitem{saphir}
J. Barth {\it et al.}, hep-ph/0307083.

\bibitem{hermes}
W. Lorenzon for the HERMES Collaboration, 
presented at the Pentaquark Workshop, Nov 6-8, 2003, 
Jefferson Lab.

\bibitem{clas2}
V. Kubarovsky, L. Guo {\it et al.}, Phys. Rev. Letts. {\bf 92}, 032001 (2004).

\bibitem{NA49}
NA49 Collaboration, hep-ex/0310014.

\bibitem{negative2}
H.G. Fischer and S. Wenig, hep-ex/0401014;
K.T. Kn\"{o}pflet {\it et al.}, 
hep-ex/0403020.

\bibitem{H1}
H1 Collaboration, hep-ex/0403017.

\bibitem{negative}
R. Cahn and G. Trilling, 
Phys. Rev. D {\bf 69}, 011501 (2004); J.Z. Bai {\it et al.}, 
[BES Collaboration], hep-exp/0402012; K.T. Kn\"{o}pflet {\it et al.}, 
hep-ex/0403020.

\bibitem{dzierba}
A.R. Dzierba {\it et al.}, hep-ph/0311125.

\bibitem{soliton}
D. Diakonov, V. Petrov, M. Polyakov, hep-ph/9703373.

\bibitem{jaffe}
R. Jaffe and F. Wilczek, hep-ph/0307341, Phys. Rev. Lett. 
{\bf 91},232003 (2003).

\bibitem{cpr}
S. Capstick, P.R. Page, and W. Roberts, Phys. Lett. B {\bf 570}, 185 (2003).

\bibitem{gk}
S.M. Gerasyuta and V.I. Kochkin, hep-ph/0310227.

\bibitem{walliser}
H. Walliser and V.B. Kopeliovich, hep-ph/0304058.

\bibitem{ma_private}
B. Wu and B.Q. Ma, Phys. Lett. B {\bf 586}, 62 (2004), hep-ph/0312326.

\bibitem{bijker}
R. Bijker, M.M. Giannini and E. Santopinto, hep-ph/0310281.

\bibitem{karliner}
M. Karliner and H.J. Lipkin, hep-ph/0402008.

\bibitem{gao}
Jefferson Lab Proposal PR04-004, Spokespersons: H. Gao and S. Nakamura.

\end{references}
\end{document}